\documentclass{article}
\usepackage{amssymb}

\input{tcilatex}
\begin{document}

\title{}
\author{}
\maketitle
\title{{\Large A Fundamental Solution to the CCC equations } {\Large \ }\\
{\Large \ \ \ \ \ \ \ \ \ }}
\author{ {\tiny \ ..\ }\ \ \ \ \ \ \ \ \ \  \ \ \ \ \ \ \ \ \ \ \ \ \ \ Ezra
T. Newman \\
%EndAName
{\tiny .\ }\ \ \ \ \ \ \ Dept of Physics and Astronomy, Univ of Pittsburgh,
\ \ \\
{\tiny .\ }\ \ \ \ \ \ \ \ \ Pittsburgh \ PA, 15260, USA}

\begin{abstract}
We display a simple solution to the Penrose CCC scenario. For this solution
we chose for the \textit{late stages} of the previous aeon a FRW, k=0,
universe with a both a cosmological constant and radiation (no mass) while
for the early stages of the 'present' aeon we have again a FRW universe,
k=0, with the same cosmological constant and again with radiation but with
mass not yet present. \ The Penrose conditions force the parameters
describing the radiation of the former and present aeons to be equal and the
transition metric in the overlap region turns out to be flat.

We further study how different rest-mass zero fields transition between the
different conformally related regions. \ These (test) fields appears to
easily allow perturbations of the geometry within the CCC scenario.
\end{abstract}

\section{Introduction}

\qquad Several years ago Penrose proposed a detailed, rather radical - and
certainly contentious - idea into cosmology. It has been referred to as
Cycle Conformal Cosmology, (CCC)\cite{Book},\cite{RandG}. \ The idea is that
there exists a non-ending sequence of aeons, each beginning with a Big-Bang
and ending (after a very long time) with an exponentially expanding
universe. \ The space-time metric, $\widehat{g}_{ab},$ of one aeon, (\textit{%
referred to as the previous aeon)}, is conformally connected to the metric, $%
\check{g}_{ab},$ of the next aeon (referred to as the\textit{\ present aeon}%
) by the conformal factor $\Omega .\ $The relationship between them is given
in the transition or overlap region via the transition metric $g_{ab},$\ by 
\begin{eqnarray}
\widehat{g}_{ab} &=&\Omega ^{2}g_{ab}  \label{ghat} \\
\check{g}_{ab} &=&\omega ^{2}g_{ab}  \label{ghacheck} \\
\omega &=&-\ \Omega ^{-1}.  \label{omega}
\end{eqnarray}

As the transition surface (or Big Bang), denoted by $\chi ,$ is approached
from the previous aeon, both $\widehat{g}_{ab}$and$\ \Omega \ $tend to
infinity so that $g_{ab}\ $is\ regular while $\omega \ $and\ $\check{g}%
_{ab}\ $vanish at the start, $\chi ,$ the Big Bang of the present aeon.

The metrics of the two aeons satisfy the Einstein equations with positive
cosmological constant scaled to be%
\begin{equation}
\Lambda =3.
\end{equation}%
The matter content of the late stages of the previous aeon and the very
early stages of the new aeon are taken to be pure radiation. (It is assumed
- a potential weak point of the theory - that all rest mass disappears in
the late stages of every aeon and returns shortly \textit{after the new Big
Bang}.)

In order to make this scenario into a predictive physical theory, Penrose
chooses a dynamical equation, with initial conditions, for the evolution of
the conformal factor $\Omega .\ $The equation chosen, the Yamabe equation,%
\begin{equation}
\square \Omega +2\varepsilon \Omega =2\Omega ^{3}  \label{yamabe}
\end{equation}%
is the special case of the transformation of the Ricci scalar under a
conformal transformation when both Ricci scalars are constant. \ $\square \ $%
is the wave operator with the transition metric, $g_{ab}$.$\ $Though $%
\varepsilon \ $is usually chosen to be \textit{one, }we will allow it to be
either \textit{one or zero}.

Penrose chooses the initial conditions for $\Omega \ $so that near $\chi ,\ $%
the norm of $\nabla _{a}\omega \ \ $(using the transition metric $g_{ab}\ $%
for the norm) is given by 
\begin{equation}
\nabla _{a}\omega \nabla ^{a}\omega =1+(Q-2)\omega ^{2}\ +0(\omega ^{3})
\label{normI}
\end{equation}%
$Q\ $is to be a given universal positive constant.

Though there are a variety of foundational theoretical questions and
fascinating real predictions and observational issues associated with this
scheme of Penrose, since it is not the thrust of this note, for completeness
and continuity of ideas we only briefly mention three of the most important:
\ 

1. The CCC scheme seems to be the first reasonable means of discussing the
so-call Big Bang of standard cosmology without invoking religion. A
well-known Astrophysicist once called me on the phone to ask "if I did not
think that the Big Bang was the proof of the existence of God." \ He even
wrote a book on the subject. In addition, the default standard cosmological
model with the inflationary scenario simply avoids the difficulties of the
Big Bang by forgetting that the difficulties exists or that it even is an
issue.

2. \ The CCC allows for an explanation of the origin of the 2$^{nd}\ \ $law
of thermodynamics.\cite{Book}

3. \ A generic CCC scenario predicts that in the CMB sky of the present aeon
there will be families of concentric circles with lower and higher than
background temperatures. \ Though it is not yet generally agreed on, there
appears to be good evidence that such families of circles do exist.\cite%
{RandG},\cite{Pawel},\cite{Pawel2} Our present simple model does not allow
for such concentric circles.

The main purpose of this note is to present a simple case of a CCC scenario
- probably the most basic or fundamental one that captures most of the
principle ideas of a CCC scenario. \ It also allows for the development of a
perturbation procedure for the development of more general CCC scenarios.

Rather than following the Penrose scheme of beginning with the metric of the
previous aeon, then solving the evolution Yamabe equation with specific
initial conditions to find the $\Omega $ and thus obtaining the present aeon
metric, we instead, chose a different procedure. \ We chose a previous aeon
model universe with metric and then \textit{guess} what the associated
present universe and metric should be. \ This allows us to construct a
conformal factor $\Omega \ \ $from the two metrics and check if the Yamabe
equation, (with initial conditions), is satisfied or if the parameters of \
the present metric (or parameters in the $\Omega $)\ must be adjusted.

More specifically we chose for both the previous and the present aeons FRW
universes with $k=0$ and cosmological constant, $\Lambda =3.\ $(We could
have chosen without much effort, k=1,-1.) The matter in both cases are pure
radiation since it is assumed in the late stages of the previous aeon (the
period of our interest) all the mass has disappeared, while for the present
aeon, in the early stages, (again the time of our interest) mass has not yet
made its appearance. \ There are only two adjustable parameters in this
scenario, the radiation parameters for each of the aeons. It turns out that
the Yamabe equation, with the initial conditions, are satisfied when the two
parameters are set equal. \ The parameter $Q$ for the initial conditions
turns out to be two.

\section{Construction of the metrics}

We begin with the conformally flat form of the $k=0$, FRW metric%
\begin{equation}
ds^{2}=R^{2}(\tau )[d\tau ^{2}-dr^{2}-r^{2}(d\theta ^{2}+\sin ^{2}\theta
d\varphi ^{2})  \label{metric}
\end{equation}%
where $\tau \ $is the conformal time (which runs from\ $\tau =0\ $to$\ \tau
=\tau _{\infty }=finite$)\ and $R(\tau )\ $is the scale factor.

[Though we do not need it, cosmic physical time is found via $dt=R(\tau
)d\tau .]$

The reduced form of the FRW differential equation, \cite{Tod}, for $\Lambda
=3,\ $pure dust and radiation, with 
\begin{equation}
S=R/R_{0},  \label{S}
\end{equation}%
($R_{0}\ $being the `present' value (arbitrary) of the scale factor) becomes%
\begin{equation}
\frac{dS}{d\tau }=R_{0}(b+aS+S^{4})^{\frac{1}{2}}.  \label{FRW}
\end{equation}

The parameters $a\ $and $b,\ $(which describe the matter and radiation
sources), are given by the ratio of the density parameters at the `present'
time,

\begin{equation}
a=\frac{\Omega _{M}}{\Omega _{\Lambda }},\ \ b=\frac{\Omega _{\gamma }}{%
\Omega _{\Lambda }}.  \label{parameters}
\end{equation}

Taking $a,$\ the mass parameter to vanish, by assumption, in the two regions
of interest, Eq.(\ref{FRW}) can be formally solved by%
\begin{equation}
\int_{S_{0}}^{S}\frac{dS}{(b+S^{4})^{\frac{1}{2}}}=R_{0}(\tau -\tau _{0}).
\label{solution I}
\end{equation}%
Though the integral can be evaluated in terms of elliptic functions, it is
far easier and more useful to evaluate it by approximations in the two
different domains - the late previous aeon where $S\ \ $tends to infinity
and the early stages of the present domain where $S\ $begins with zero.

To keep the notation for the two aeons clear and separate the variables of
the previous aeon will be written with a $hat,\ $e.g., $\widehat{S},\widehat{%
b},\ $etc.$,$while for the present the same variables will have an inverted
hat$,(hachek)\ $e.g.,\ $\check{S},\check{b}.$

\subsection{Previous Aeon}

By rewriting Eq. (\ref{solution I}) as

\begin{equation}
\int_{\widehat{S}_{0}}^{\widehat{S}}\frac{d\widehat{S}}{\widehat{S}^{2}(1+%
\widehat{b}\widehat{S}^{-4})^{\frac{1}{2}}}=\widehat{R}_{0}(\widehat{\tau }-%
\widehat{\tau }_{0})  \label{solution II}
\end{equation}%
and expanding the denominator for large $S,\ $we have%
\begin{equation}
\int_{S_{0}}^{S}d\widehat{S}(\widehat{S}^{-2}-\frac{1}{2}b\widehat{S}^{-6})=%
\widehat{R}_{0}(\widehat{\tau }-\widehat{\tau }_{0}).
\end{equation}%
This is easily integrated as%
\begin{equation}
-\widehat{S}^{-1}+\frac{1}{10}\widehat{b}\widehat{S}^{-5}+\widehat{S}%
_{0}^{-1}-\frac{1}{10}b\widehat{S}_{0}^{-5}=\widehat{R}_{0}(\widehat{\tau }-%
\widehat{\tau }_{0}).
\end{equation}

Choosing $\widehat{S}_{0}\ $as infinity and $\widehat{\tau }_{0}=\widehat{%
\tau }_{\infty },\ $we have%
\begin{equation}
\widehat{S}^{-1}-\frac{1}{10}\widehat{b}\widehat{S}^{-5}=-\widehat{R}_{0}(%
\widehat{\tau }-\widehat{\tau }_{\infty }).
\end{equation}

Finally, defining 
\begin{equation}
\Gamma =\widehat{\tau }-\widehat{\tau }_{\infty },  \label{Gamma}
\end{equation}%
$($which is negative since$\ \widehat{\tau }<\widehat{\tau }_{\infty })\ $\
and iterating for small $\Gamma ,\ ($starting from $\widehat{S}%
_{start}^{-1}=-\widehat{R}_{0}\Gamma \ )\ $we have our solution near $\Gamma
=0:$%
\begin{eqnarray}
\widehat{S}^{-1} &=&-\widehat{R}_{0}\Gamma +\frac{1}{10}\widehat{b}S^{-5} \\
\widehat{S}_{start}^{-1} &=&-\widehat{R}_{0}\Gamma  \\
\widehat{S}^{-1} &=&-\widehat{R}_{0}\Gamma (1+\frac{1}{10}\widehat{b}%
\widehat{R}_{0}^{4}\Gamma ^{4}).
\end{eqnarray}%
Using $\widehat{R}\equiv \widehat{S}\widehat{R}_{0}\ $and inverting, we
finally have our scale function for the late stages of the previous aeon.%
\begin{eqnarray}
\widehat{R} &\equiv &-\frac{\Gamma ^{-1}}{(1+\frac{1}{10}\widehat{b}\widehat{%
R}_{0}^{4}\Gamma ^{4})},  \label{Rhat} \\
&=&-\Gamma ^{-1}(1-\frac{1}{10}\widehat{b}\widehat{R}_{0}^{4}\Gamma ^{4}). 
\nonumber
\end{eqnarray}

\subsection{Present Aeon}

Returning to the integral, Eq.(\ref{solution I}), and applying it to the
very early stages of the present aeon, i.e., for small $\check{S},\ \ $the
integral is approximated by%
\begin{eqnarray}
\int_{\check{S}_{0}}^{\check{S}}\frac{d\check{S}}{\sqrt{\check{b}}(1+\frac{%
S^{4}}{\check{b}})^{\frac{1}{2}}} &=&\check{R}_{0}(\check{\tau}-\check{\tau}%
_{0}), \\
\int_{\check{S}_{0}}^{\check{S}}d\check{S}(1-\frac{1}{2}\frac{\check{S}^{4}}{%
\check{b}}) &=&\sqrt{\check{b}}\check{R}_{0}(\check{\tau}-\check{\tau}_{0}).
\end{eqnarray}

Performing the integration and using $\check{S}_{0}=0\ $at $\check{\tau}%
_{0}=0,\ $followed by the iteration, leads to%
\begin{equation}
\check{S}=\sqrt{\check{b}}\check{R}_{0}\check{\tau}+\frac{1}{10}b^{3/2}%
\check{R}_{0}^{5}\check{\tau}^{5}.
\end{equation}

Using $\check{R}=\check{S}\check{R}_{0},$with the definition, $\Gamma =%
\check{\tau},\ $gives us the scale factor for the present aeon,%
\begin{equation}
\check{R}=\sqrt{\check{b}}\check{R}_{0}^{2}\Gamma +\frac{1}{10}b^{3/2}\check{%
R}_{0}^{6}\Gamma ^{5}  \label{Rhacheck}
\end{equation}

The conformal time parameter $\Gamma ,\ $which is positive for the present
aeon is simply the continuation of the negative valued $\Gamma \ $of the
previous aeon. $\Gamma =0\ $is\ the\ time of the transition or Big Bang.

\subsection{The Conformal Factor and the Metrics}

Introducing the parameters $\widehat{J}\ $and $\check{J}\ \ $by\ 
\begin{equation}
\widehat{J}=\sqrt{\widehat{b}}\widehat{R}_{0}^{2}\ \ \text{and\ \ }\check{J}=%
\sqrt{\check{b}}\check{R}_{0}^{2}  \label{J}
\end{equation}%
the two scale factors of the past and present aeon are%
\begin{eqnarray}
\widehat{R} &=&-\Gamma ^{-1}(1-\frac{1}{10}\widehat{J}^{2}\Gamma ^{4}),
\label{the R's} \\
\check{R} &=&\check{J}\Gamma (1+\frac{1}{10}\check{J}^{2}\Gamma ^{4}). 
\nonumber
\end{eqnarray}

The associated metrics are thus%
\begin{eqnarray}
\widehat{g}_{ab} &=&\widehat{R}^{2}\eta _{ab},  \label{g's} \\
\check{g}_{ab} &=&\check{R}^{2}\eta _{ab}.  \nonumber
\end{eqnarray}

By eliminating $g_{ab}\ \ $from the two relationships, Eqs.(\ref{ghat}) and (%
\ref{ghacheck}), with (\ref{omega}), we obtain

\begin{equation}
\widehat{g}_{ab}=\Omega ^{4}\check{g}_{ab},  \label{OMEGA^4}
\end{equation}%
so that, from (\ref{g's}),%
\[
\Omega ^{4}=\frac{\widehat{R}^{2}}{\check{R}^{2}} 
\]%
or%
\begin{eqnarray}
\Omega ^{2} &=&-\frac{\widehat{R}}{\check{R}}=\frac{\Gamma ^{-2}(1-\frac{1}{%
10}\widehat{J}^{2}\Gamma ^{4})}{\check{J}\Gamma (1+\frac{1}{10}\check{J}%
^{2}\Gamma ^{4})},  \label{OMEGAsq} \\
&\approx &\check{J}^{-1}\Gamma ^{-2}(1-\frac{1}{10}\widehat{J}^{2}\Gamma
^{4})(1-\frac{1}{10}\check{J}^{2}\Gamma ^{4}),  \nonumber \\
&\approx &\check{J}^{-1}\Gamma ^{-2}(1-\frac{1}{10}[\check{J}^{2}+\widehat{J}%
^{2}]\Gamma ^{4}).  \nonumber
\end{eqnarray}

Our conjecture is that this $\Omega ,\ $perhaps with the adjustment of the
parameters, ($\widehat{J},\check{J}),$ satisfies the Penrose conditions for
the transition from the previous aeon to the present one.

\qquad The following are a series of approximate relationships easily
derived from the $\Omega \ $that are useful for checking the Penrose
transition conditions.

\begin{eqnarray}
~%
%TCIMACRO{\U{3a9} }%
%BeginExpansion
\Omega
%EndExpansion
^{-2} &=&-\frac{\check{R}}{\widehat{R}}=\check{J}\Gamma ^{2}(1+\frac{1}{10}(%
\check{J}^{2}+\widehat{J}^{2})\Gamma ^{4},  \label{1} \\
%TCIMACRO{\U{3a9} }%
%BeginExpansion
\Omega
%EndExpansion
&=&-\check{J}^{-1/2}\Gamma ^{-1}(1-\frac{1}{20}(\check{J}^{2}+\widehat{J}%
^{2})\Gamma ^{4},  \label{2} \\
~d%
%TCIMACRO{\U{3a9} }%
%BeginExpansion
\Omega
%EndExpansion
&=&\check{J}^{-1/2}\{\Gamma ^{-2}+\frac{3}{20}(\check{J}^{2}+\widehat{J}%
^{2})\Gamma ^{2}\}d\Gamma ,  \label{3} \\
~\omega &=&-%
%TCIMACRO{\U{3a9} }%
%BeginExpansion
\Omega
%EndExpansion
^{-1}=\check{J}^{1/2}\Gamma (1+\frac{1}{20}(\check{J}^{2}+\widehat{J}%
^{2})\Gamma ^{4},  \label{4} \\
\Gamma &=&~\check{J}^{-1/2}\omega ,  \label{5} \\
g_{ab} &=&\Omega ^{-2}\widehat{g}_{ab}=-\widehat{R}\check{R}\eta
_{ab}=R^{2}\eta _{ab},  \label{6} \\
\sqrt{-g} &=&\sqrt{-\det g_{ab}}=R^{4}  \label{6a} \\
R^{2} &=&\mathfrak{-}\widehat{R}\check{R}=\check{J}(1+\frac{1}{10}[\check{J}%
^{2}-\widehat{J}^{2}]\Gamma ^{4}),  \label{7} \\
R &=&\check{J}^{1/2}(1+\frac{1}{20}[\check{J}^{2}-\widehat{J}^{2}]\Gamma
^{4}).  \label{8}
\end{eqnarray}

Notice that the intermediate metric $g_{ab\ }$is flat when 
\begin{equation}
\check{J}=\widehat{J}.  \label{equal J}
\end{equation}

\section{The Penrose Transition Conditions}

The most important of the Penrose conditions\cite{RandG} on $\Omega \ $is
the Yamabe equation

\begin{equation}
\square \Omega +2\varepsilon \Omega =2\Omega ^{3}.  \label{Yamabe.II}
\end{equation}%
It becomes, using the transition metric\ and $\Gamma =x^{0},\ \partial
_{\Gamma }F=F^{\prime },$

\begin{eqnarray}
(-g)^{-1/2}\partial _{a}((-g)^{1/2}g^{ab}\partial _{b}\Omega )+(\varepsilon
2)\Omega &=&2\Omega ^{3},  \label{yamabeA} \\
\eta ^{qb}R^{2}\partial _{a}(\partial _{b}\Omega )+\eta ^{ab}\partial
_{a}R^{2}\partial _{b}\Omega +R^{4}(\varepsilon 2)\Omega &=&2R^{4}\Omega
^{3},  \label{B} \\
R^{2}\Omega ^{\prime \prime }+(R^{2})^{\prime }\Omega ^{\prime
}+R^{4}(\varepsilon 2)\Omega &=&2R^{4}\Omega ^{3}.  \label{C}
\end{eqnarray}%
Inserting Eqs.(\ref{7})\ and (\ref{2}) into (\ref{C}), leads after a lengthy
calculation to%
\[
\lbrack \check{J}^{2}-\widehat{J}^{2}]\Gamma =0 
\]%
or the condition that 
\begin{equation}
\check{J}=\widehat{J}.  \label{equalJ}
\end{equation}%
\qquad \qquad

With this equality we have that the transition metric $g_{ab\ }$is flat in
the neighborhood of the transition surface and the Yamabe equation is
satisfied.

To examine the remaining Penrose conditions, namely the initial conditions
for the Yamabe equation, we must find $\nabla _{a}\omega \ $and its norm, $%
\nabla _{a}\omega \nabla ^{a}\omega .\ $They are calculated with the
transition metric, using the equality of the two $J^{\prime }$s \ (their
decorations are discarded):

\begin{eqnarray}
\nabla _{a}\omega &=&J^{\frac{1}{2}}(1+\frac{1}{2}J^{2}\Gamma ^{4})\delta
_{a}^{0},  \label{grad&norm} \\
\nabla _{a}\omega \nabla ^{a}\omega &=&1+J^{2}\Gamma ^{4}.  \nonumber
\end{eqnarray}%
\qquad

Using (\ref{5}). i.e., using $\omega \ $as the 'time" parameter, the norm
becomes%
\begin{equation}
\nabla _{a}\omega \nabla ^{a}\omega =1+\omega ^{4}.
\end{equation}%
Comparing this with the Penrose condition, Eq.(\ref{normI}), we see that
indeed, as required, the norm at $\Gamma =0\ $is one and that $Q=2.$

\subsection{Miscellaneous}

Penrose discusses\cite{RandG} several other relationships derived from the
conformal factor $\Omega .\ $For completeness, we briefly examine several of
them, comparing the results from our model with those of Penrose\cite{RandG}.

1. There is the one form $\Pi \ $and its norm in the transition region given
by Penrose as 
\begin{eqnarray*}
\Pi &=&\frac{d%
%TCIMACRO{\U{3a9} }%
%BeginExpansion
\Omega
%EndExpansion
}{%
%TCIMACRO{\U{3a9} }%
%BeginExpansion
\Omega
%EndExpansion
^{2}-1}=\frac{d\omega }{1-\omega ^{2}}, \\
\Pi ^{a}\Pi _{a} &=&1+Q\omega ^{2}+O(\omega ^{3}),
\end{eqnarray*}%
which become in our model%
\begin{eqnarray*}
\Pi &=&J^{1/2}(1+J\Gamma ^{2}+\frac{3}{2}J^{2}\Gamma ^{4})d\Gamma , \\
\Pi ^{a}\Pi _{a} &=&1+2J\Gamma ^{2}+4J^{2}\Gamma ^{4}.
\end{eqnarray*}%
In the norm, if we change the coordinate from $\Gamma \ $to $\omega $, by
Eq.(\ref{5})$,$the norm becomes 
\[
\Pi ^{a}\Pi _{a}=1+2\omega ^{2}+O(\omega ^{4}), 
\]%
yielding again that $Q=2.$

2. \ Penrose has the divergence of $\Pi ^{a}\ $as\cite{RandG}%
\[
\nabla _{a}\Pi ^{a}=2Q\omega +O(\omega ^{2}), 
\]%
while we have it as%
\[
\nabla _{a}\Pi ^{a}=2\omega +O(\omega ^{3}). 
\]

There is thus a disagreement, by the factor of 2, when taking $Q=2$. \ This
disagreement probably\cite{tod 2} has its origin with the fact that in the
Yamabe equation we take $\varepsilon $\ to be zero while Penrose takes it to
be one. It appears likely that if we had chosen for the FRW universes, the
curvature $k=1$ case rather than our choice $k=0,\ $this disagreement would
vanish.

\subsection{Test Fields}

As a prelude to studying perturbation on our simple CCC scenario, we
describe the transformations of both Maxwell fields and linearized Weyl
tensor fields under the conformal transformations that relate the three
metrics, $\widehat{g}_{ab},\ g_{ab},\ \check{g}_{ab}.$

Since there are variety of different conformal transformations that we must
consider, we first use generic variables, with generic decorations, $%
(\clubsuit ,\blacklozenge ),$ and a generic conformal factor, $\widetilde{%
%TCIMACRO{\U{3a9} }%
%BeginExpansion
\Omega
%EndExpansion
}$, to describe the transformations. Afterwards they are specialized to the
ones considered in the CCC discussion.

%comment
There are a large variety of conformal rescaling, including their inverses,
e.g., from Minkowski space to a FRW space, or from one aeon to another with
their inverses, that we must deal with. \ Considerable care must be used in
going between these different cases. 

For the metrics we use, 
\begin{equation}
g_{ab}^{\clubsuit }=\widetilde{%
%TCIMACRO{\U{3a9} }%
%BeginExpansion
\Omega
%EndExpansion
}^{2}g_{ab}^{\blacklozenge },  \label{g'}
\end{equation}%
for the Maxwell fields, 
\begin{equation}
F_{ab}^{\clubsuit }\ =F_{ab}^{\blacklozenge },  \label{F'}
\end{equation}%
for the Weyl tensor components, 
\begin{equation}
C_{\ \ \ \ bcd}^{\clubsuit a}=C_{\ \ \ \ b\ cd}^{\blacklozenge a},
\label{C"}
\end{equation}%
and for the NP tetrad fields, $\ $%
\begin{eqnarray}
l^{\clubsuit a} &=&\widetilde{%
%TCIMACRO{\U{3a9} }%
%BeginExpansion
\Omega
%EndExpansion
}^{-2}l^{\blacklozenge a},\ \ \ \ \ \ \ l_{a}^{\clubsuit
}=l_{a}^{\blacklozenge },\   \label{tetrad} \\
n^{\clubsuit a} &=&n^{\blacklozenge a},\ \ \ \ \ \ \ \ \ \ \ \
n_{a}^{\clubsuit }=\widetilde{%
%TCIMACRO{\U{3a9} }%
%BeginExpansion
\Omega
%EndExpansion
}^{2}n_{a}^{\blacklozenge },  \nonumber \\
m^{\clubsuit a} &=&\widetilde{%
%TCIMACRO{\U{3a9} }%
%BeginExpansion
\Omega
%EndExpansion
}^{-1}m^{\blacklozenge a},\ \ \ \ \ m_{a}^{\clubsuit }=\widetilde{%
%TCIMACRO{\U{3a9} }%
%BeginExpansion
\Omega
%EndExpansion
}m_{a}^{\blacklozenge },  \nonumber \\
\overline{m}^{\clubsuit a} &=&\widetilde{%
%TCIMACRO{\U{3a9} }%
%BeginExpansion
\Omega
%EndExpansion
}^{-1}\overline{m}^{\blacklozenge a},\ \ \ \ \ \overline{m}_{a}^{\clubsuit }=%
\widetilde{%
%TCIMACRO{\U{3a9} }%
%BeginExpansion
\Omega
%EndExpansion
}m_{a}^{\blacklozenge }.  \nonumber
\end{eqnarray}

Using Eq.(\ref{tetrad}) with (\ref{F'}) and (\ref{C"}), one easily
calculates the spin-coefficient versions of the transformations. For the
Maxwell case we have

\begin{eqnarray}
\phi _{0}^{\clubsuit } &=&\widetilde{%
%TCIMACRO{\U{3a9} }%
%BeginExpansion
\Omega
%EndExpansion
}^{-3}\phi _{0}^{\blacklozenge }  \label{phi's} \\
\phi _{1}^{\clubsuit } &=&\widetilde{%
%TCIMACRO{\U{3a9} }%
%BeginExpansion
\Omega
%EndExpansion
}_{1}^{-2}\phi ^{\blacklozenge }  \nonumber \\
\ \ \ \ \ \phi _{2}^{\clubsuit } &=&\widetilde{%
%TCIMACRO{\U{3a9} }%
%BeginExpansion
\Omega
%EndExpansion
}^{-1}\phi _{2}^{\blacklozenge },  \nonumber
\end{eqnarray}%
with%
\begin{equation}
\phi _{0}^{\blacklozenge }\equiv l^{\blacklozenge a}m^{\blacklozenge
b}F_{ab}^{\blacklozenge },\ \ \ \phi _{1}^{\blacklozenge }\equiv 
%TCIMACRO{\U{bd}}%
%BeginExpansion
{\frac12}%
%EndExpansion
(l^{\blacklozenge a}n^{\blacklozenge b}+m^{\blacklozenge a}\overline{m}%
^{\blacklozenge b})F_{ab}^{\blacklozenge },\ \ \ \phi _{2}^{\blacklozenge
}\equiv \ \overline{m}^{\blacklozenge a}n^{\blacklozenge
b}F_{ab}^{\blacklozenge },  \label{phi}
\end{equation}%
while for the linearized Weyl tensor we have \qquad \qquad 

\begin{eqnarray}
\Psi _{0}^{\clubsuit } &=&\widetilde{%
%TCIMACRO{\U{3a9} }%
%BeginExpansion
\Omega
%EndExpansion
}^{-4}\Psi _{0}^{\blacklozenge },  \label{PSI's} \\
\Psi _{1}^{\clubsuit } &=&\widetilde{%
%TCIMACRO{\U{3a9} }%
%BeginExpansion
\Omega
%EndExpansion
}^{-3}\Psi _{1}^{\blacklozenge },  \nonumber \\
\Psi _{2}^{\clubsuit } &=&\widetilde{%
%TCIMACRO{\U{3a9} }%
%BeginExpansion
\Omega
%EndExpansion
}^{-2}\Psi _{2}^{\blacklozenge },  \nonumber \\
\Psi _{3}^{\clubsuit } &=&\widetilde{%
%TCIMACRO{\U{3a9} }%
%BeginExpansion
\Omega
%EndExpansion
}^{-1}\Psi _{3}^{\blacklozenge },  \nonumber \\
\Psi _{4}^{\clubsuit } &=&\Psi _{4}^{\blacklozenge }.  \nonumber
\end{eqnarray}%
with%
\begin{eqnarray}
\Psi _{0}^{\blacklozenge } &\equiv &-l^{\blacklozenge a}m^{\blacklozenge
b}l^{\blacklozenge c}m^{\blacklozenge d}C_{abcd}^{\blacklozenge },
\label{PSI} \\
\Psi _{1}^{\blacklozenge } &\equiv &-l^{\blacklozenge a}n^{\blacklozenge
b}l^{\blacklozenge c}m^{\blacklozenge d}C_{abcd}^{\blacklozenge },  \nonumber
\\
\Psi _{2}^{\blacklozenge } &\equiv &-l^{\blacklozenge a}m^{\blacklozenge b}%
\overline{m}^{\blacklozenge c}n^{\blacklozenge d}C_{abcd}^{\blacklozenge }, 
\nonumber \\
\Psi _{3}^{\blacklozenge } &\equiv &-l^{\blacklozenge a}n^{\blacklozenge b}%
\overline{m}^{\blacklozenge c}n^{\blacklozenge d}C_{abcd}^{\blacklozenge }, 
\nonumber \\
\Psi _{4}^{\blacklozenge } &\equiv &-\overline{m}^{\blacklozenge
a}n^{\blacklozenge b}\overline{m}^{\blacklozenge c}n^{\blacklozenge
d}C_{abcd}^{\blacklozenge }.  \nonumber
\end{eqnarray}

Our first application of these relations is to the transition from a
solution of either Maxwell's equations or the "flat-space" Bianchi
Identities for the Weyl tensor in Minkowski space,$\ $to the conformally
related solutions in a FRW space-time. \ 

In particular, we consider transitioning the flat-space solutions$\ $to\ our
previous aeon solutions.

The decorations then change from $Y^{\blacklozenge }\ $to$\ Y\ ^{flat}$ and\ 
$Y^{\clubsuit }$\ to $\widehat{Y},$ with $\widetilde{%
%TCIMACRO{\U{3a9} }%
%BeginExpansion
\Omega
%EndExpansion
}\ $becoming $\widehat{R}=-\Gamma ^{-1}(1-\frac{1}{10}\widehat{b}\widehat{R}%
_{0}^{4}\Gamma ^{4}).\ $

Close to the Big Bang surface, $\chi ,$ $\Gamma \approx 0,\ $ so for
simplicity we take $\widehat{R}=-\Gamma ^{-1},\ $ with $\Gamma \ $negative.

A flat-space Maxwell field ($\phi _{0}^{flat},\phi _{1}^{flat},\phi
_{2}^{flat}$) or linear Weyl tensor field ($\Psi _{0}^{flat},\Psi
_{1}^{flat},\Psi _{2}^{flat},\Psi _{3}^{flat},\Psi _{4}^{flat}$) becomes
from (\ref{phi's}) and (\ref{PSI's})

\begin{eqnarray}
\widehat{\phi }_{0} &=&-\Gamma ^{3}\phi _{0}^{flat},  \label{phihat} \\
\widehat{\phi }_{1} &=&\Gamma ^{2}\phi _{1}^{flat},  \nonumber \\
\widehat{\phi }_{2} &=&-\Gamma \phi _{2}^{flat},  \nonumber
\end{eqnarray}%
and the corresponding Weyl fields%
\begin{eqnarray}
\widehat{\Psi }_{0} &=&\Gamma ^{4}\Psi _{0}^{flat},  \label{PSIhat} \\
\widehat{\Psi }_{1} &=&-\Gamma ^{3}\Psi _{1}^{flat},  \nonumber \\
\widehat{\Psi }_{2} &=&\Gamma ^{2}\Psi _{2}^{flat},  \nonumber \\
\widehat{\Psi }_{3} &=&-\Gamma \Psi _{3}^{flat},  \nonumber \\
\widehat{\Psi }_{4} &=&\Psi _{4}^{flat}.  \nonumber
\end{eqnarray}%
\qquad 

The Maxwell fields and almost all Weyl tensor components, \textit{vanish in
the previous aeon} as $\chi \ $is approached.

However in the transition region where we conformally rescale via Eq.(\ref%
{ghat}), $g_{ab}=\Omega ^{-2}\widehat{g}_{ab},\ $with (comparing with Eq.(%
\ref{g'})), $\widetilde{%
%TCIMACRO{\U{3a9} }%
%BeginExpansion
\Omega
%EndExpansion
}=%
%TCIMACRO{\U{3a9} }%
%BeginExpansion
\Omega
%EndExpansion
^{-1}=-J^{1/2}\Gamma (1+\frac{1}{10}J^{2}\Gamma ^{4}),\ $or approximately $%
\widetilde{%
%TCIMACRO{\U{3a9} }%
%BeginExpansion
\Omega
%EndExpansion
}=$ $%
%TCIMACRO{\U{3a9} }%
%BeginExpansion
\Omega
%EndExpansion
^{-1}\approx -J^{1/2}\Gamma ,\ $the Maxwell and Weyl fields \textit{return
to their flat-space values}, modified by the numerical factor $\sqrt{J}.\ $%
For example, from $\ \phi _{2}^{\clubsuit }=\widetilde{%
%TCIMACRO{\U{3a9} }%
%BeginExpansion
\Omega
%EndExpansion
}_{2}^{-1}\phi ^{\blacklozenge },\ $we have that 
\begin{eqnarray}
\phi _{2} &=&\widetilde{%
%TCIMACRO{\U{3a9} }%
%BeginExpansion
\Omega
%EndExpansion
}^{-1}\widehat{\phi }_{2}=-J^{-1/2}\Gamma ^{-1}\widehat{\phi }_{2},
\label{phi2} \\
&=&J_{2}^{-1/2}\phi ^{flat}.  \nonumber
\end{eqnarray}

We thus have the result that in the transition region the two massless
fields go thru the Big Bang smoothly.

Finally to see how these same fields behave near the Big Bang in the present
aeon we use Eq.(\ref{OMEGA^4}),\ i.e.$,\ \widehat{g}_{ab}=\Omega ^{4}\check{g%
}_{ab},$and compare it with Eq.(\ref{g'}), so that $\widetilde{%
%TCIMACRO{\U{3a9} }%
%BeginExpansion
\Omega
%EndExpansion
}=\Omega ^{-2}\approx J^{2}\Gamma ^{2}.\ $Using this with (\ref{phi's}) and (%
\ref{PSI's}), along with (\ref{phihat}) and (\ref{PSIhat}), the test fields
are found in the present aeon to be%
\begin{eqnarray}
\check{\phi}_{0} &=&J^{-6}\Gamma _{0}^{-3}\phi _{0}^{flat},
\label{phihacheck} \\
\check{\phi}_{1} &=&J^{-4}\Gamma ^{-2}\phi _{1}^{flat},  \nonumber \\
\check{\phi}_{2} &=&J^{-2}\Gamma ^{-1}\phi _{2}^{flat},  \nonumber
\end{eqnarray}%
and the corresponding Weyl fields%
\begin{eqnarray}
\check{\Psi}_{0} &=&J^{-8}\Gamma ^{-4}\Psi _{0}^{flat},  \label{PSIhacheck}
\\
\check{\Psi}_{1} &=&J^{-6}\Gamma ^{-3}\Psi _{1}^{flat},  \nonumber \\
\check{\Psi}_{2} &=&J^{-4}\Gamma ^{-2}\Psi _{2}^{flat},  \nonumber \\
\check{\Psi}_{3} &=&J^{-2}\Gamma ^{-1}\Psi _{3}^{flat},  \nonumber \\
\check{\Psi}_{4} &=&\Psi _{4}^{flat}.  \nonumber
\end{eqnarray}

\textit{All the fields in the present aeon are singular at }$\chi \ $and
diminish$\ $as $\Gamma \ $increases from its zero value.

If we consider any Maxwell or Weyl field (or any rest-mass zero field) in
the conformally related flat space of the previous aeon and refer to it as $%
\Phi ^{flat},\ $it and all its conformally related counterparts in the
different regions can be taken as functions of the flat-space null
coordinate system $(u,r,\theta ,\phi )$. \ We thus have, generically
speaking,

\begin{equation}
\Phi =\Phi (u,\ r,\ \theta ,\ \phi ).  \label{generic}
\end{equation}

Replacing $u,\ $via$\ u=$ $\tau -r\ $and $\Gamma =\tau -\tau _{\infty }$, (%
\ref{Gamma}) by 
\begin{equation}
u=\Gamma +\tau _{\infty }-r  \label{u&Gamma}
\end{equation}%
we have%
\begin{equation}
\Phi =\Phi (\Gamma +\tau _{\infty }-r,\ r,\ \theta ,\ \phi ).  \label{CAPPHI}
\end{equation}

If, as a special case, we consider $\Phi \ $to have support only on a
light-cone, (e.g., a violent event at $\tau =\tau ^{\#}\ $in\ the\ previous
aeon or perhaps even a series of them following close to each other) chosen
as $u=\tau ^{\#},\ $the support of the field lies on 
\begin{equation}
r=\Gamma +\tau _{\infty }-\tau ^{\#}.  \label{r=}
\end{equation}%
In the limit $\Gamma =0,\ $($\chi ,\ $the cross-over or Big Bang surface)
the field 
\begin{equation}
\Phi =\Phi (\tau ^{\#},\ \tau _{\infty }-\tau ^{\#},\ \theta ,\ \phi )
\label{CAPPHIatr}
\end{equation}%
has support only on the sphere

\begin{equation}
r=r_{\infty }=\tau _{\infty }-\tau ^{\#},  \label{rinfinity}
\end{equation}%
which is finite and a known function of the initial conditions for the
dynamics of the previous aeon.

%comment
As an aside we remark that if this field on $\chi \ $or rather shortly
afterwards (i.e., last scattering surface) were to scatter by interactions
with other fields in the present aeon, the effect could potentially be seen
by present observers looking back on their past-light cones. These past
cones intersect the sphere $r=\tau _{\infty }-\tau ^{\#}\ $in a circle (or
for a series of events, a concentric family of circles) thus potentially
appearing to an observer as a circle (or a series of circles) on the CMB
sky. \ 

\section{Discussion}

Working essentially with the Penrose CCC scenario, we have chosen a special
simple case to explore - a previous aeon late time FRW universe with
radiation that transitions to a present aeon early FRW universe also with
radiation. \ This case appeared to us as a natural example of a CCC (with
perhaps some tweaking) that must exist almost \textit{a priori}.\ There were
two needed adjustments for this to work. One was that the intermediate
metric had to have a vanishing cosmological constant. This was manifested in
the choice, in the Yamabe equation, that $\varepsilon =0\ $rather than
Penrose$^{\prime }$s preference, $\varepsilon =1.\ $The other adjustment was
that the two radiation parameters must coincide. A by-product of these
adjustments is that the intermediate metric is flat.

We next considered the behavior of different (test) rest-mass fields
(Maxwell and linear Weyl tensor fields) as they go thru the different
conformally related regions.

Our intent is to take this simple CCC case and consider the test fields as
sources for perturbations on the CCC background. The most obvious thing to
do is to treat both Maxwell and linearized Weyl tensor fields that have
their support mainly on a light-cone from some earlier origin of the
previous aeon (as we have done) and follow the perturbed geometry from the
past thru to the present aeon.

\qquad This hopefully will yield detailed examples of greater complexity,
for analysis in the CCC scenario.

\section{Acknowledgments}

We would like to thank our colleagues Paul Tod and Pawel Nurowski for their
help and insight. But above all we thank Roger Penrose for his patience in
helping us go thru and understand the difficulties of this fascinating new
view of cosmology.

\bigskip


\begin{thebibliography}{99}
\bibitem{Book} R. Penrose, \textit{Cycles of Time: An Extraordinary New View
of the Universe}.

(Bodley Head, London 2010) ISBN 978-0-224-08036-1.

\bibitem{RandG} V. G. Gurzadyan and R. Penrose, \textit{On CCC-Predicted
Concentric low-variance Circles in the CMB Sky}, Eur.

Phys. J. Plus (2013) 128, 22-38.

\bibitem{Pawel} Krzysztof A. Meissner, Pawel Nurowski and Blazej Ruszczycki,
\ \textit{Structures in the Microwave Background Radiation},

Proc. R. Soc. A 2013 469, 20130116, 15 May 2013

\bibitem{Pawel2} Daniel An, Krzysztof A. Meissner, Pawel Nurowski, \textit{%
Structures in the Planck map of the CMB}, arXiv:1307.5737

\bibitem{Tod} K.P. Tod, \textit{Penrose's Circles in the CMB and a Test of
Inflation}, General Relativity and Gravitation, Volume 44, Issue 11, \qquad
\qquad 

pp.2933-2938

\bibitem{tod 2} Private conversations with K. P. Tod
\end{thebibliography}
\end{document}